\documentclass[prc,twocolumn,superscriptaddress,aps]{revtex4-1}
\usepackage{amsmath,amssymb,epsfig,bm,mathrsfs,feynmp,color}
\usepackage{slashed}
\newcommand{\be}{\begin{equation}}
\newcommand{\ee}{\end{equation}}
\newcommand{\bea}{\begin{eqnarray}}
\newcommand{\eea}{\end{eqnarray}}

\newcommand{\de}{\partial}

\newcommand{\vecq}{{\mathbf q}}
\newcommand{\vecp}{{\mathbf p}}
\newcommand{\veck}{{\mathbf k}}

\newcommand{\vecalpha}{\bm \alpha}
\newcommand{\vecgamma}{\bm \gamma}

\newcommand{\vectau}{{\bm \tau}}
\newcommand{\vecpi}{{\bm \pi}}
\newcommand{\ie}{{\it i.e.}}

\definecolor{red}{rgb}{0.8,0,0}
\definecolor{orange}{rgb}{0.8,0.2,0.0}
\definecolor{blue}{rgb}{0.3,0.0,0.8}

\begin{document}

\title{Leading-order nucleon self-energy in relativistic chiral effective field theory}

 \author{Giuseppe Colucci}
\affiliation{Institute for Theoretical Physics, 
J. W.\ Goethe University, Max-von-Laue-Str.\ 1, \\
D-60438 Frankfurt am Main, Germany}

\author{Armen Sedrakian} 
\affiliation{Institute for Theoretical Physics, 
J. W.\ Goethe University, Max-von-Laue-Str.\ 1, \\
D-60438 Frankfurt am Main, Germany}

\author{Dirk H. Rischke} 
\affiliation{Institute for Theoretical Physics, 
J.  W.\ Goethe University, Max-von-Laue-Str.\ 1, \\
D-60438 Frankfurt am Main, Germany}

\affiliation{Frankfurt Institute for Advanced Studies, 
Ruth-Moufang-Str.\ 1,\\
D-60438 Frankfurt am Main, Germany}


\begin{abstract}
  We apply thermal field theory methods to compute microscopically the
  nucleon self-energy arising from one-pion exchange in
  isospin-symmetric nuclear matter and neutron matter.  A
  self-consistent numerical scheme is introduced and its convergence
  is demonstrated. The repulsive contribution from the Fock exchange
  diagram to the energy per nucleon in symmetric nuclear matter is
  obtained.
\end{abstract}

\maketitle
\section{Introduction}
\label{sec:1}

The equation of state of bulk nuclear matter has attracted
considerable attention over time, as it has a substantial impact on
the properties of neutron stars as well as finite nuclei.  The
equation of state is determined by two, generally related,
ingredients: the force between nucleons and the many-body
approximation used to compute the thermodynamic properties of nuclear
matter. Nuclear interactions are accurately modeled in terms of
potentials arising from meson exchange: on the one hand models have
been used which are based on phase-shift equivalent one-boson
exchange~\cite{1995PhRvC..51...38W,*Stoks:1994wp,*Machleidt:2000ge,
  *Lacombe:1980dr} and, on the other hand, more recently,
(multi)-pion-exchange potentials based on chiral power
counting~\cite{Weinberg:1990rz,*Weinberg:1991um,*PhysRevC.53.2086,*2011PhR...503....1M,*2009RvMP...81.1773E}.
The diversity of ab-initio many-body approaches includes variational
Monte Carlo
methods~\cite{1998PhRvC..58.1804A,*1993RvMP...65..817B,*2007PhRvL..98j2503G,*MNL2:MNL2829}
and propagator methods with (medium-) renormalized soft interactions
as, e.g., in non-relativistic and covariant Brueckner-type
theories~\cite{PhysRevC.32.1057,*PhysRevC.84.024301,*PhysRevC.84.035801,*Sedrakian:2006mq,*2011arXiv1111.0695S}. Perturbative
calculations of nuclear matter properties were revived with the advent
of soft chiral potentials~\cite{PhysRevC.83.031301,*PhysRevC.82.014314}.

In this work we apply thermal field theory (TFT) methods to compute
the leading-order nucleon self-energy in isospin-symmetric nuclear and neutron
matter. The method has been applied in the past in the context of QED
and QCD plasmas to compute the quasiparticle energies of electrons and
quarks, respectively, in a thermal
medium~\cite{1993PhRvD..48.1390B,*2001PhRvD..63f5003B}. In the
low-energy regime of interest to us, the relevant degrees of freedom
are nucleons and pions. Chiral symmetry is an (approximate) symmetry
of the strong interaction. Because chiral symmetry is spontaneously
broken in nature, pions emerge as (pseudo-) Goldstone bosons of the
theory. The description of the strong interaction in terms of chiral
Lagrangians admits certain approximation schemes in terms of power
counting of small
quantities~\cite{1979PhyA...96..327W,1996NuPhB.478..629K}, which in
principle allow for a systematic order-by-order improvement of a given
calculation. Below, we will combine relativistic TFT methods and the
chiral Lagrangian description of nuclear interactions to address the
computation of the nucleon self-energy.

Approaches similar to ours were previously developed in
Refs.~\cite{2004PhRvC..69c5211F,2000PhLB..474....7L,2002NuPhA.697..255K,*2002NuPhA.700..343K,*2012PrPNP..67..299W,2011AnPhy.326..241L,*2010JPhG...37a5106O}.
Fraga et al.~\cite{2004PhRvC..69c5211F} derived analytical expressions
for the zero-temperature self-energy of a nucleon due to pion exchange
in dilute nuclear matter to leading order in the chiral expansion. Lutz et
al.~\cite{2000PhLB..474....7L} and Kaiser et
al.~\cite{2002NuPhA.697..255K,*2002NuPhA.700..343K,*2012PrPNP..67..299W}
(hereafter KFW)
used chiral Lagrangians and expansions in small Fermi momentum to
construct equations of state in the heavy-baryon limit. The saturation
in Ref.~\cite{2000PhLB..474....7L} arises due to correlations induced
by the one-pion-exchange interaction. In
Ref.~\cite{2002NuPhA.697..255K,*2002NuPhA.700..343K,*2012PrPNP..67..299W}
two-pion exchange produces nuclear binding at the three-loop level with a
suitably adjusted momentum cut-off.  An effective field
  theory of nuclear matter with nucleons and pions, which allows for
  both local as well as pion-mediated multi-nucleon interactions, was
  developed in Ref.~\cite{2011AnPhy.326..241L,*2010JPhG...37a5106O}
  in the heavy-baryon limit 
  and the main trends for the energy density of symmetric nuclear and
  neutron matter were already reproduced at next-to-leading order in
  their power-counting scheme. 

  In applying TFT to nuclear matter in a relativistic setting, we have
  to maintain the covariance of the theory by using fully relativistic
  thermal propagators for nucleons and pions.  Keeping the Lorentz
  symmetries intact is of advantage, in particular, for computing the
  scattering and radiation amplitudes with full propagators and
  renormalized vertices. Specifically the computation
  of particular electro-weak processes in nuclear and neutron matter 
  (or transitions in nuclei), which are driven by currents with
  particular Lorentz symmetry are conveniently carried out in the Dirac
  basis. 

The second goal of our work is to maintain
  self-consistency among the propagators and the self-energies of the
  theory, which means that the iterations are performed until the
  Schwinger-Dyson equation for the nucleons is fulfilled. If a firm
  perturbative expansion (with power-counting rules) exists, a
  self-consistent approach generates higher-order terms in every
  iteration, and therefore is not really necessary.  Nevertheless,
  using the leading-order term in the chiral expansion of the
  Lagrangian, we will check the impact of these higher-order terms in
  a self-consistent solution.  

  This paper is structured as follows. In Sec.~\ref{sec:lagrangian} we
  discuss the chiral Lagrangian. Section~\ref{sec:thermalQFT} uses TFT
  to compute the pion contribution to the nucleon self-energy. Our
  numerical method and results for the self-energy are presented in
  Sec.~\ref{sec:results}. Our conclusions are collected in
  Sec.~\ref{sec:conclusions}. We use natural units $\hbar = c = k_B =
  1$. Four-vectors are denoted with capital letters, for instance
  $P^\mu=(p^0, \vecp)$.

\section{Lagrangians}
\label{sec:lagrangian}
\begin{figure}[t]
 \centering
\includegraphics[width=8.5cm,height=3.cm]{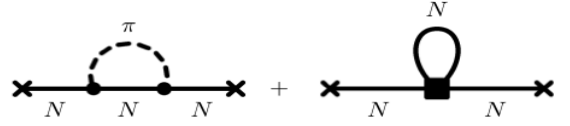}
 \caption{
The diagrams contributing to the self-energy of the
nucleon to lowest order.
The first one represents the chiral one-pion exchange contribution
to the nucleon self-energy. Solid lines refer to nucleons, dashed
lines to pions and dots to pion-nucleon vertices. The
second diagram is the contribution from the four-fermion contact
interaction, with the square vertex representing the two-body
scattering matrix. 
}
 \label{fig:fock_diagram}
\end{figure}

Low-energy nuclear dynamics can be constructed on the basis of the
pion and nucleon degrees of freedom starting from a chiral
Lagrangian.  The interaction Lagrangian $ {\cal L}_{\pi N}$ between
nucleons and pions is constructed such as to reflect the spontaneous chiral
symmetry breaking of strong interactions at low energies.  Since the
interactions of Goldstone bosons must vanish at zero-momentum transfer
and in the chiral limit (i.e., the pion mass $m_\pi\rightarrow 0$), 
a low-energy expansion in powers (the so-called chiral dimension) of
the ratio of the momentum or the pion mass over ($4 \pi$ times) 
the pion decay constant can be performed.
Consequently, the Lagrangian can be written as 
\be\label{eq:ChiExp1} 
{\cal L}_{\pi N}
= 
{\cal L}_{\pi N}^{(1)} + {\cal L}_{\pi
  N}^{(2)} 
+ \ldots , 
\ee 
where the superscript labels the order of the chiral
dimension.  The terms in the expansion (\ref{eq:ChiExp1}) are
constructed by introducing the following $SU(2)$ matrix $U$ in flavor
space 
\be 
U = \exp\left(i\frac{\vectau\cdot\vecpi}{f_\pi}\right) = 1 +
\frac{i}{f_\pi} \mbox{\boldmath $\tau$} \cdot \mbox{\boldmath $\pi$}
-\frac{1}{2f_\pi^2} \mbox{\boldmath $\pi$}^2 + \ldots, 
\ee where
$\vectau$ is vector of Pauli matrices in isospin space, $\vecpi$ is the
isotriplet of pions, and $f_\pi$ the pion decay constant.  The 
leading-order term is given by \cite{1988NuPhB.307..763G}
\be\label{eq:lo_lagrangian} 
{\cal L}^{(1)}_{\pi N} = \bar{\psi}
\left(i\gamma^\mu D_\mu - m + \frac{g_A}{2} \gamma^\mu \gamma_5
  u_\mu\right) \psi \, , 
\ee 
where $\psi$ is the nucleon field,
$\bar\psi = \psi^\dagger\gamma_0$, $m$ is the nucleon mass, and $g_A$ is
the axial-vector coupling. The physical value of $g_A$ is determined
from neutron beta decay and is given by $g_A =
1.2695\pm0.0029$. $D_\mu$ is the covariant derivative,
\be 
D_\mu = \de_\mu + \Gamma_\mu 
\ee 
where $\Gamma_\mu$ is the so-called
chiral connection which couples an even number of pions to the
nucleon and is defined as 
\bea 
\Gamma_\mu & = &
i\{\xi^\dagger, \partial_\mu \xi\}
= i (\xi^\dagger \partial_\mu \xi + \xi \partial_\mu \xi^\dagger)
\nonumber \\
& = & - \frac{1}{4f_\pi^2} \, \vectau \cdot (\vecpi \times\de_\mu
\vecpi) + \ldots 
\label{eq:WT}
\eea with $\xi=\sqrt{U}$.  The Lagrangian (\ref{eq:lo_lagrangian})
also includes the axial-vector current $u_\mu$ which couples an odd
number of pions to the nucleon \be u_\mu = i[\xi^\dagger, \partial_\mu
\xi] = i (\xi^\dagger \partial_\mu \xi - \xi \partial_\mu \xi^\dagger)
= - \frac{1}{f_\pi} \, \mbox{\boldmath $\tau$} \cdot \partial_\mu
\mbox{\boldmath $\pi$} + \ldots \; .  \ee Keeping only the
lowest-order term in the chiral Lagrangian (\ref{eq:lo_lagrangian})
the pion-nucleon Lagrangian reads \cite{2011PhR...503....1M}
\be\label{eq:1piexchange} {\cal L}_{\pi
  N}^{(1)}=\bar\psi\left(i\gamma^\mu\de_\mu- m
  -\frac{g_A}{2f_\pi}\gamma^\mu\gamma_5\vectau\cdot\de_\mu\vecpi\right)\psi,
\ee where we have neglected the Weinberg-Tomozawa contribution arising
from the chiral connection (\ref{eq:WT}).  The chiral one-pion
interaction term in Eq.~(\ref{eq:1piexchange}) takes the light
dynamical degrees of freedom, \ie, pions, explicitly into account.
The complete interaction Lagrangian includes  to
lowest order  an additional
four-fermion contact term. Thus, the Lagrangian of the system
can be written as \
\be \label{eq:full} {\cal L} = {\cal L}_{\rm free} +{\cal L}_{\pi
  N}^{(1)}+ {\cal L}_{NN},
\ee 
where the first term is the Lagrangian of the non-interacting system
and the last term corresponds to the four-fermion
interaction. Figure~\ref{fig:fock_diagram} shows the two distinct leading-order 
contributions to the self-energy of a nucleon
originating from the tree-level vertices in the Lagrangian
(\ref{eq:full}).

\section{Pion contribution to the self-energy}
\label{sec:thermalQFT}
\subsection{Leading-order contribution}

In this section we evaluate the one-pion contribution to the
nucleon self-energy, corresponding to the left diagram in Fig.\ 
\ref{fig:fock_diagram}, within the imaginary-time
formalism. The free covariant propagator of the nucleons in
energy-momentum space is given by
\be\label{fermion_prop}
{S}_0(K)=
\frac{\Lambda^+_k\gamma_0}{ik_n + \mu - E_k} 
+ \frac{\Lambda^-_k\gamma_0}{ik_n + \mu + E_k},
\ee
where the zeroth component of the four-momentum $K^\mu\equiv (k^0,\veck)$
takes discrete values, 
$k^0=ik_{n} = (2n+1)\pi i T $, $n\in \mathbb{Z}$ and $T$ is the
temperature. $\Lambda^\pm_k$ are projectors onto positive $(+)$ and
negative $(-)$ energy states,
\be
\Lambda^\pm_k = \frac{1}{2 E_k}\left[E_k\pm 
(\vecalpha\cdot{\veck}+m\gamma_0)\right],
\ee
where 
 $E_k^2={\veck}^2 + m^2$ is the dispersion relation for
non-interacting nucleons,  $m$ is their mass, 
 and   $\vecalpha \equiv \gamma_0 \vecgamma$. 
The free pion propagator is given by
\be\label{boson_prop}
 D_0(Q)
=\frac{1}{2\omega_q}\left(\frac{1}{i\omega_n-\omega_q}
-\frac{1}{i\omega_n+\omega_q}\right)\;,
\ee
where the zeroth component of the four-momentum $Q^\mu\equiv
(q^0,\vecq)$ takes discrete values $q^0=i\omega_{n} = 2n\pi i T $,
$n\in \mathbb{Z}$, and $\omega_q^2=q^2 + m_\pi^2$ is the 
dispersion relation for non-interacting pions,
with $m_\pi$ being the free pion mass. In terms of the free
propagators (\ref{fermion_prop}) and (\ref{boson_prop}) the 
one-pion exchange contribution to the nucleon
self-energy reads
\bea\label{generic_self-energy}
\Sigma(P) & = &
-\frac{3 g_A^2}{4 f_\pi^2} T\int\!\!\frac{d^3\veck}{(2\pi)^3}
\sum_{ik_n}D_0(P-K)
\gamma_5(\slashed P-\slashed K)\nonumber\\
&\times&S_0(K)\gamma_5(\slashed
P-\slashed K)\; ,
\eea 
where $P^\mu=(p^0,\vecp)\equiv (ip_n,\vecp)$.
We substitute
the propagators (\ref{fermion_prop}) and (\ref{boson_prop}) 
into Eq.\ (\ref{generic_self-energy}) and carry out the
summation over the fermionic Matsubara frequency $ik_n$. In general,
this sum generates physically distinct processes involving all
possible combinations of bosons and fermions and their antiparticles.
The result of the summation can be arranged according to these underlying
processes, but it is more convenient to separate the self-energy into
the ``vacuum'' and ``thermal'' parts $\Sigma(P) =
\Sigma_0(P) + \Sigma_T(P)$, where the
vacuum part is given by 
\bea\label{t0self}
\Sigma_0(P) & = & \frac{3 g_A^2}{4 f_\pi^2}
\int\frac{d^3\veck}{(2\pi)^3}\frac{1}{2\omega_{pk}}\nonumber\\
&\times&\Biggl[\frac{\gamma_5(\slashed P-\slashed K)\Lambda^+_k\gamma_0\gamma_5
(\slashed P-\slashed K)}{ip_n + \mu - E_k -\omega_{pk}}
\bigg|_{k_0=ip_n-\omega_{pk}}\nonumber\\
    &+&\frac{\gamma_5(\slashed P-\slashed K)\Lambda^-_k\gamma_0\gamma_5
(\slashed P-\slashed K)}{ip_n + \mu + E_k +\omega_{pk}}
\bigg|_{k_0=E_k + \mu}\Biggr]\;,
\eea
and the thermal part is given by 
\begin{widetext}
\bea\label{finite_t_self}
\nonumber\Sigma_T(P) & = & \frac{3 g_A^2}{4 f_\pi^2}
\int\frac{d^3\veck}{(2\pi)^3}\frac{1}{2\omega_{pk}}\\
\nonumber& \bigg\{  & \left[\frac{1}{ip_n + \mu - E_k -\omega_{pk}}
-\frac{1}{ip_n + \mu - E_k +\omega_{pk}}\right]\gamma_5
(\slashed P-\slashed K)\Lambda^+_k\gamma_0\gamma_5(\slashed P-\slashed K)
\big|_{k_0=E_k - \mu}n_F(E_k-\mu)\\
\nonumber& - & \left[\frac{1}{ip_n + \mu + E_k -\omega_{pk}}
-\frac{1}{ ip_n + \mu + E_k +\omega_{pk}}\right]\gamma_5(\slashed P-\slashed K)
\Lambda^-_k\gamma_0\gamma_5(\slashed P-\slashed K)\big|_{k_0=- E_k -
  \mu} n_F(E_k-\mu)\\
\nonumber& - & \gamma_5(\slashed P-\slashed K)
\left[\frac{\Lambda^+_k\gamma_0}{
ip_n + \mu - E_k -\omega_{pk}}+\frac{\Lambda^-_k\gamma_0}{ip_n + \mu + 
E_k -\omega_{pk}}\right]\gamma_5(\slashed P-\slashed K)
\big|_{k_0=ip_n - \omega_{pk}}n_B(\omega_{pk})\\
& - & \gamma_5(\slashed P-\slashed K)\left[\frac{\Lambda^+_k\gamma_0}{
    ip_n 
+ \mu - E_k +\omega_{pk}}+\frac{\Lambda^-_k\gamma_0}{ip_n + \mu + E_k 
+\omega_{pk}}\right]\gamma_5(\slashed P-\slashed K)
\big|_{k_0=ip_n + \omega_{pk}}n_B(\omega_{pk})\bigg\}\;,
\eea
where $\omega_{pk}^2 = (\vecp - \veck)^2 + m_\pi^2$ and
$n_{F/B} (x) = [\exp(x/T) \pm 1]^{-1}$ are the Fermi/Bose
distribution functions.
The retarded self-energy is obtained by analytical continuation,
\ie,  $ip_n\rightarrow p_0 +
i0^+$.  We have verified that in the case of a Yukawa interaction
Eq.~(\ref{finite_t_self}) transforms to the well-known expression for
the self-energy of a fermion in finite-temperature quantum field
theory \cite{1993PhRvD..48.1390B,Kapusta-book}.

At sufficiently low temperature the occupation number of anti-particles is
so small that we can neglect their
contribution in Eq.\ (\ref{finite_t_self}). Furthermore, we assume that
there is no macroscopic occupation of pionic modes in nuclear matter
at any temperature and density of interest; therefore, we also drop
the contributions proportional to the bosonic occupation numbers. 
The remaining contribution arises from the pole at $k_0 \equiv E_k -
\mu$ and we arrive at
\bea
\label{eq:sigma_approx}
\Sigma_T(P) & = &  \frac{3 g_A^2}{4 f_\pi^2} 
\int\frac{d^3\veck}{(2\pi)^3}\frac{1}{2\omega_{pk}} 
\left[\frac{1}{p_0 - k_0 - \omega_{pk} + i0^+} - \frac{1}{p_0 - k_0 +
    \omega_{pk} + i0^+}\right](s+\slashed Q)n_F(k_0)\;,
\eea
where we have defined the quantities 
$s = - (m/2E_k)(P-K)^2$ and $Q^\mu = (q_0,\vecq)$, with components
\bea
\label{0_lambda}
q_0    & = & \frac{1}{2}[(p_0-k_0)^2+(\vecp-\veck)^2] -
\frac{1}{E_k}(\vecp-\veck)
\cdot\veck(p_0 - k_0)\;,\\
\label{vector_lambda}
\vecq & = & -\frac{1}{2E_k}\left[(P-K)^2\veck +
  2(\vecp-\veck)\cdot\veck(\vecp-\veck)\right] +
(p_0-k_0)(\vecp-\veck)\;.  
\eea 
Later on we will enforce self-consistency in evaluating the self-energy.
This requires a Lorentz decomposition of the self-energy,
which in the most general case is given by
\be\label{full_lorentz_structure}
\Sigma(P) = \Sigma_s(P) + \gamma_5\Sigma_{ps}(P) +
\gamma^\mu\Sigma_\mu(P) + \gamma_5\gamma^\mu\Sigma^A_\mu(P) +
\sigma^{\mu\nu}\Sigma_{\mu\nu}(P)\;.
\ee
The requirements of parity conservation, translational and rotational
invariance, as well as time-reversal invariance, reduce this most
general decomposition to the following form 
\be\label{lorentz_structure}
\Sigma(P) = \Sigma_s(P) + \gamma_0\Sigma_0(P) 
+ {\mbox{\boldmath $\gamma$}}\cdot{\mathbf p}\Sigma_v(P)\;.
\ee
Equation (\ref{eq:sigma_approx}) can now be projected onto its 
Lorentz components by multiplying it with $1, \gamma_0$, and 
$\vecgamma$, and taking the trace
over the $\gamma$-matrices. Keeping only the thermal part of the
self-energy (and dropping the index $T$ on the self-energies), 
this leads us to the following decomposition coefficients
\bea\label{sigmas}
\Sigma_s (P)& = & -\frac{3g_A^2}{4 f_\pi^2} 
\int\frac{d^3\veck}{(2\pi)^3}\frac{1}{2\omega_{pk}}
 \left[\frac{1}{p_0 - k_0 - \omega_{pk} + i0^+} - \frac{1}{p_0 - k_0 
+ \omega_{pk} + i0^+}\right] n_F(k_0)  \frac{m}{2E_k}(P-K)^2\;,\\
\Sigma_0 (P)& = & 
        + \frac{3g_A^2}{4 f_\pi^2} \int\frac{d^3\veck}{(2\pi)^3}
        \frac{1}{2\omega_{pk}} \left[\frac{1}{p_0 - k_0 - \omega_{pk}
            + i0^+} 
- \frac{1}{p_0 - k_0 + \omega_{pk} + i0^+}\right]
n_F(k_0)\nonumber \\
        &  \times  & \left[\frac{1}{2}((p_0-k_0)^2+({\vecp}-\veck)^2) +
          \frac{1}{E_k}(\vecp-\veck)\cdot\veck(p_0 - k_0)\right]\;,
\label{sigma0}\\
\nonumber
\vert \vecp\vert \Sigma_v (P)& = & 
        + \frac{3 g_A^2}{4 f_\pi^2} \int\frac{d^3\veck}{(2\pi)^3}
        \frac{1}{2\omega_{pk}} \left[\frac{1}{p_0 - k_0 
- \omega_{pk} + i0^+} - \frac{1}{p_0 - k_0 + \omega_{pk} + i0^+}
\right]n_F(k_0)\\
 &  \times  & \Biggl\{\frac{1}{2E_k}\left[(P-K)^2\veck 
+2(\vecp-\veck)\cdot\veck(\vecp-\veck)\right] 
- (p_0-k_0)(\vecp-\veck)\Biggr\}\cdot\hat \vecp\;,\label{sigmav}
\eea
where $\hat\vecp = \vecp/\vert\vecp\vert$ is a unit vector.
Equations (\ref{sigmas})--(\ref{sigmav}) are our final result for the
chiral one-pion-exchange contribution to the nucleon self-energy. It is evident that
the other terms in Eq.~(\ref{finite_t_self}), which could become
important at higher temperatures and densities, can be evaluated in a
completely analogous way.
\end{widetext}

\subsection{Further approximations}

For numerical computations the factor
$s+\slashed Q$ in Eq.~(\ref{eq:sigma_approx}) 
can be simplified. 
We start by rewriting the expression 
\bea
2E_k\slashed Q
& = & 2E_k(q_0\gamma_0 - \vecq\cdot\vecgamma) 
=  \mu\left[(p_0-k_0)^2+(\vecp - \veck)^2\right]\gamma_0 \nonumber\\
&+& 2(P-K)\cdot K\slashed P- (P^2-K^2)\slashed K \nonumber\\
&-& 2\mu(p_0-k_0)(\vecp-\veck)\cdot\vecgamma\;.
\eea
For the densities and temperatures of interest, 
nucleons are constrained to the vicinity of 
their Fermi surface, therefore the momentum of the nucleon
can be expressed as $\mathbf p = p_F\hat{\mathbf n} + \delta\mathbf
p$. Here, $p_F$ is
the nucleon Fermi momentum, 
$\hat{\mathbf n}$ is a unit vector, and $\delta\mathbf p$ is
the residual momentum, with $\vert \delta \mathbf p\vert \ll p_F$.
Furthermore, the relativity parameter $x = p_F/m$ is small as well; 
numerically, we have
\be 
x \approx 0.28 \left(\frac{n}{n_0}\right)^{1/3}\; ,
\ee
where $n$ is the density of the systems, 
$n_0 = 0.16\;{\rm fm}^{-3}$ is the nuclear saturation density.
Then, the nucleon energy 
is $E_p \approx m(1+\vecp^2/2m^2)$ and the chemical potential 
$\mu \approx m(1+p_F^2/2m^2)$.
This implies that $p_0 = E_p -
\mu \approx (\vecp^2-p_F^2)/2m\approx x\, 
\hat{\mathbf n}\cdot \delta \mathbf p 
$ 
is small compared to $\vert
\vecp\vert \approx p_F$. Therefore, we can
replace $(P-K)^2\approx-(\mathbf p - \mathbf k)^2$. 

With these approximations we obtain
\bea
2E_k~\slashed Q & \approx & 2m~\slashed Q 
 \approx   \mu(\vecp-\veck)^2\gamma_0 
-2(\vecp-\veck)\cdot\veck \slashed P \nonumber\\
&+& (\vecp^2-\veck^2)\slashed K -(\vecp^2-\veck^2)(\vecp-\veck)\cdot \vecgamma\nonumber\\
& \approx & (\vecp-\veck)^2[(p_0 + \mu)\gamma_0 - \vecp\cdot\vecgamma]
\eea
where we used the fact that 
$p_0-k_0\approx (\vecp^2-p_F^2)/2m-
(\veck^2-p_F^2)/2m=(\vecp^2-\veck^2)/2m$.

Therefore, the factor $s+\slashed Q$ in the integrand of the self-energy 
(\ref{eq:sigma_approx}) reads now 
\be 
s+\slashed Q \approx
\frac{(\mathbf p-\mathbf k)^2}{2m}[m + (p_0 + \mu)\gamma_0 - \vecgamma
\cdot \vecp] \;.
\ee 

Separating the contributions from 
particles and anti-particles in the boson propagator, with these
approximations the self-energy reads
\bea
\Sigma(P) & \approx & [m + (p_0 + \mu)\gamma_0 - \vecgamma\cdot \vecp]
\left[\sigma_+(P)+\sigma_-(P) \right]\nonumber\\
& \equiv & [m + (p_0 + \mu)\gamma_0 - \vecgamma\cdot \vecp]~
\sigma(P)\label{eq:sigma_+}\;,
\eea
where 
\be\label{eq:sigma_pm}
\sigma_{\pm}(P)
 = \pm  \frac{3 g_A^2}{8 m f_\pi^2} 
\int\frac{d^3\veck}{(2\pi)^3} \frac{1}{2\omega_{pk}}
\frac{(\mathbf p - \mathbf k)^2}{p_0 - k_0 \mp \omega_{pk} + i\eta}
n_F(k_0),
\ee
is the {\it reduced self-energy}.
Therefore, the coefficients of the Lorentz decomposition of the 
self-energy can be expressed in terms of $\sigma(p)$,
\bea\label{eq:coefficients_sigma}
&&\Sigma_s(P) \approx m\sigma(P),
\quad\Sigma_0(P) \approx (p_0 + \mu)\sigma(P),\nonumber\\
&&\Sigma_v(P) \approx - \sigma(P)\;.
\eea
The real part of the self-energy can now be computed with 
the help of the Dirac identity. We obtain
by explicitly evaluating the Cauchy principal value of 
Eq.~(\ref{eq:sigma_pm})
\bea\label{eq:re_sigma}
{\rm Re}[\sigma(P)] & \equiv & {\rm Re}[\sigma_+(P) + \sigma_-(P)] \nonumber\\
&=&\frac{3 g_A^2}{ 32 m \pi^2 f_\pi^2}
\int_0^\infty\!\!\! d\vert\veck\vert\,
\veck^2\,
\frac{n_F(k_0)}{2\vert \vecp\vert\vert \veck\vert}~
\int_{-1}^1\!\! dx\frac{(\vecp - \veck)^2}{x-x_0}\nonumber\\
\eea
where $x_0 = 
[\vecp^2 + \veck^2 + m_\pi^2 -  
(p_0 - k_0)^2]/(2\vert \vecp\vert\vert \veck\vert) $. 

\subsection{Enforcing self-consistency}

The self-consistency of the numerical computation of the
self-energy is achieved by replacing the free
nucleon propagator in Eq.~(\ref{generic_self-energy}) by the full
nucleon propagator, determined by the Schwinger-Dyson equation
\bea
S^{-1}(P) & = & S_0^{-1}(P) - \Sigma(P)
                 =   [(p_0 + \mu)\gamma_0 - \vecgamma
\cdot \vecp]\nonumber\\&\times &[1 -
                {\rm Re}\sigma(P)] 
- m[1 + {\rm   Re}\sigma(P)] + {\rm Im }\Sigma(P),\nonumber\\
\eea
where the free nucleon propagator is given by 
$
S_0^{-1}(P) =  (p_0 + \mu)\gamma_0 - \vecgamma\cdot\vecp - m
$
and we have used Eq.\ (\ref{eq:sigma_+}) for the nucleon self-energy.
The roots of ${\rm det} S^{-1}(P)$ determine the excitation
spectrum of the system,
\be\label{qpa_spectrum}
p_0^* = E_p^* - \mu^*\;,
\ee
where 
\bea\label{E_eff}
E_p^* & = & \sqrt{{\vecp^*}^2 + {m^*}^2}\;, \\
\label{m_eff}
m^*      & = & m[1 + {\rm Re}\sigma(P^*)]\;,\\
\label{p_eff}
\vecp^* & = & \vecp[1 - {\rm Re}\sigma(P^*)]\;,\\
\label{mu_eff}
\mu^* & = & \mu[1 - {\rm Re}\sigma(P^*)]\;.  \eea We achieve
self-consistency for the self-energy by replacing the free quantities
$m$, $\vecp$, $\veck$, and $\mu$ in the integrand (but not in the
integration measure) of Eq.\ (\ref{eq:re_sigma}) by the corresponding
renormalized $m^*$, $\vecp^*$, $\veck^*$, and $\mu^*$.  In practice,
we start by computing (\ref{eq:re_sigma}) with the free quantities,
which define the renormalized quantities (\ref{E_eff})--(\ref{mu_eff})
to first order in the iteration process.  We repeat the previous step
until convergence is reached.  Approximations made in the numerical
part of this work are purely technical and concern only the
computation of traces in the self-energy diagram. It is evident from
Eqs.~(\ref{E_eff})--(\ref{mu_eff}) that the pole structure and,
therefore, the quasiparticle spectra as well as the Lorentz structure
of the self-energies remains fully relativistic. During the iteration
procedure the quantities in Eqs.  (\ref{E_eff})--(\ref{mu_eff}) are
corrected in each iteration differently according to the Lorentz
structure of the self-energy and propagators. Thus the Dirac structure
of the self-energies and propagators is maintained.

\section{Results}
\label{sec:results}

We will not attempt to evaluate the contributions to the energy of
nuclear matter from diagrams other than the one-pion exchange
discussed in the previous section. For that purpose we define the
energy per nucleon, excluding its rest mass $m$, via the formula
\be\label{energypernucleon} 
\frac{E}{N} = {\cal T} + {\cal U} - m\;, 
\ee 
where ${\cal T}$ is the kinetic energy, ${\cal U}$ is the potential
energy and $N$ is the particle number. In terms of the renormalized
quantities (\ref{m_eff})--(\ref{mu_eff}) the kinetic energy reads
\be \label{eq:T}
{\cal T} = \frac{g_\tau}{\pi^2n}\int d|\vecp |\, \vecp^2 
\left(\frac{m^*}{E_p^*}m + \frac{|\vecp^*|}{
    E_p^*}|\vecp|\right)n_F(E_p^*-\mu^*)\;, 
\ee 
where $g_{\tau}$ is the isospin
degeneracy factor ($g_{\tau} = 2$ in isospin-symmetric nuclear matter and 
$g_{\tau} =1$ in neutron matter).  The potential energy
${\cal U}$ is given by 
\bea\label{eq:V} 
{\cal U } &=& \frac{g_\tau}{2\pi^2n}\int d|\vecp|\, \vecp^2
\left[\frac{m^*}{E_p^*} m + E_p - \frac{|\vecp^*|}{E_p^*}|\vecp|
\right]\nonumber\\
&\times&\sigma(P^*)
n_F(E_p^*-\mu^*)\;. 
\eea 
As we consider only spin-unpolarized matter, the
spin summation has been carried out in Eqs.\ (\ref{eq:T}) and
(\ref{eq:V}).


\subsection{Symmetric nuclear matter}

The Fock diagram evaluated in the previous section represents chiral
one-pion exchange between two nucleons.  Chiral power counting
requires to leading order an additional two-body contact term. The
corresponding self-energy diagrams are shown by the left and the right
diagrams in Fig.~\ref{fig:fock_diagram}.  In passing we note that if
in the nuclear medium the contact is density independent then its
contribution is given simply by $\Sigma_c = t^{(2)}n,$ where $n$ is
the density and $t^{(2)}$ is the two-body contact.

\begin{figure}[tb]
\centering
\includegraphics[width=8cm]{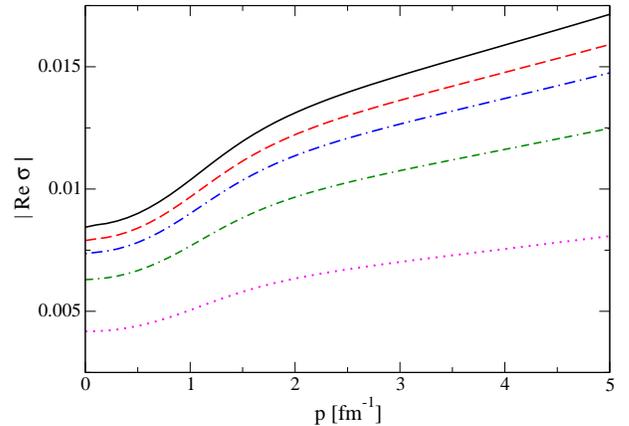}
\caption{ (Color online)
Real part of the reduced nucleon self-energy as a function of momentum $p$
at $n=0.1\;{\rm fm}^{-3}$ and $T=0$.
The dotted (violet) line is the lowest-order result and simultaneously
the starting point of the iteration. The full (black) line represents
the final self-consistent result. The other
lines correspond to 2,3,4,5 (from bottom to top) iteration steps.
}
\label{fig:iterations}
\end{figure}
  The reduced self-energy due to the chiral one-pion exchange in
  isospin-symmetric nuclear matter is shown in
  Fig.~\ref{fig:iterations} at $T=0$ and $n=0.1\;{\rm fm}^{-3}$.  As
  one observes, the self-consistency procedure converges rapidly. In
  order to achieve a relative accuracy $\le 10^{-6}$ one needs about
  10 iterations, the exact number of iteration depending on the
  density. For small densities the number of required iterations is
  small, while for large densities the number of iterations needed to
  achieve convergence is larger. While the self-energy does not tend
  to zero for large external momenta, the result for the energy of the
  system is convergent due to an additional integration over a Fermi
  distribution function.  Our numerical implementation of the
  self-consistency shows that the perturbative expansion is not
  perfect and iterations are necessary (see Fig.~\ref{fig:iterations}).
  We see that chiral power counting at non-zero density and
  temperature is not reliable anymore, since several new scales
  (temperature and chemical potential) appear.

\begin{figure}[tb]
\begin{center}
\includegraphics[angle=0,width=8cm]{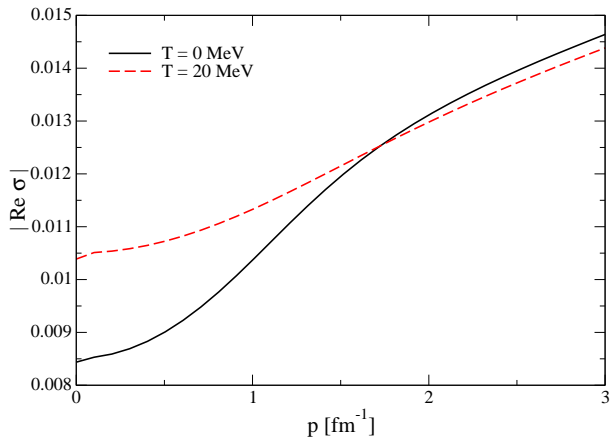}
\caption{ (Color online)
Dependence of the one-pion exchange contribution to the nucleon 
self-energy in isospin-symmetric nuclear matter on the 
momentum at density $n = 0.1$ fm$^{-3}$. 
The solid line shows the real part of the self-energy 
for temperature $T=0$ and the dashed line for $T=20$ MeV.
}
\label{fig:self1}
\end{center}
\end{figure}

  In Fig.\ \ref{fig:self1} we show the reduced self-energy at zero
  temperature in comparison to its form at $T=20$ MeV. The change with
  temperature can be seen to be rather moderate. In the non-zero
  temperature case the self-energy is larger for low momenta than at
  zero temperature, while for higher momenta the variation of
  temperature does not affect the result very much. This is due to the
  fact that we deal with comparatively low temperatures, therefore the
  temperature is a relevant scale only at low momenta.

\begin{figure}[tb]
\centering
\includegraphics[width=8cm]{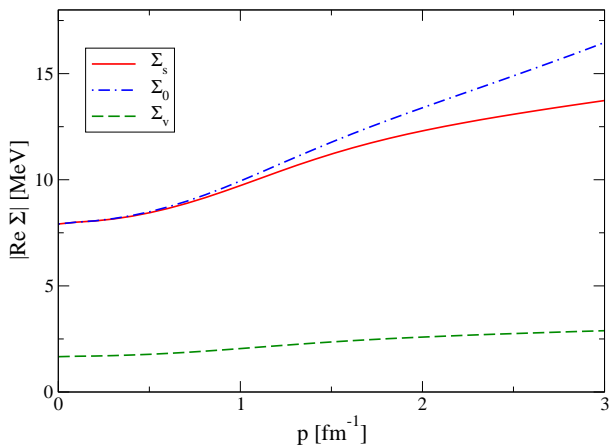}
\caption{ (Color online)
The Lorentz components of the real part of the nucleon 
self-energy as a function of momentum $p$
at $n=0.1\;{\rm fm}^{-3}$ and $T=0$.
}
\label{fig:SE_components}
\end{figure}
In Fig.\ \ref{fig:SE_components} we show the Lorentz components of the
real part of the full self-energy at $n=0.1$ fm$^{-3}$ and $T=0$.  It
can be seen that the vector self-energy is substantially smaller than
the other contributions to the self-energy, as expected. 

\begin{figure}[tb]
\centering
\vskip 1cm
\includegraphics[width=8cm]{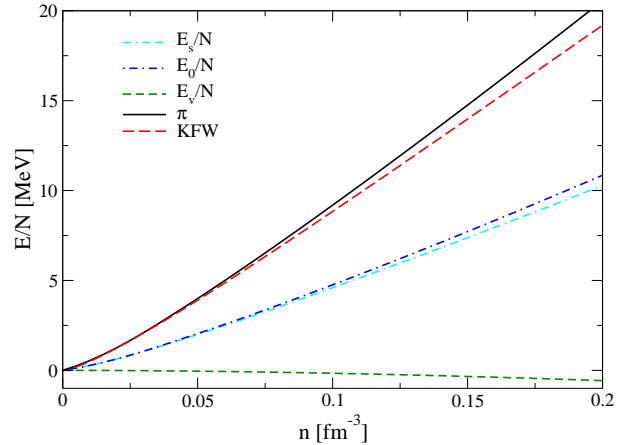}
\caption{ (Color online) Dependence of various contributions to the
  energy per particle on density. The full result is shown by the solid
  (black) line, the KFW result is shown by the dashed (red) line. The
  remaining lines show the energy per particle according 
  to the contribution of various Lorentz
  components: the scalar component $E_s$ (double dash-dotted, cyan), 
the  zero component $E_0$ (dash-dotted, blue)  and
  vector component $E_v$ (dashed, green). }
\label{fig:DeltaE}
\end{figure}

Our numerical calculations of the Fock contribution to the self-energy
of nucleons can be validated in certain limiting cases.  We verified that
our numerical result for the real part of the reduced self-energy is
in good agreement with the analytical result quoted in Ref.\
\cite{2004PhRvC..69c5211F} at zero temperature. A further test is the
comparison of the energies per particle of nuclear matter with those
of KFW, where the energy per particle was computed directly without a
reference to the self-energy.  
Their result for the zero-temperature one-pion-exchange  Fock
diagram, which is an expansion in the small relativity parameter $x$,
reads~\cite{2002NuPhA.697..255K,*2002NuPhA.700..343K,*2012PrPNP..67..299W}
\begin{eqnarray}  \label{eq:KFW1}
E{(T=0)} = {3g_A^2 p_F^3\over 16\pi^2
    f_\pi^2}\left(a_0 +  a_2x^2\right)\, ,\end{eqnarray}
where the coefficients of this expansion  $a_0$ and $a_2$ are functions of the ratio
$p_F/m_\pi$ alone and can be calculated analytically~\cite{INSET}.
In Fig.\ \ref{fig:DeltaE} we show the
contributions of the one-pion exchange to the energy of the matter
from various Lorentz components along with the full result which is
the sum of the three components. Our full result is in  good
agreement with the analytical expression 
by KFW, Eq.~(\ref{eq:KFW1}),  also shown in Fig.\
\ref{fig:DeltaE}. 

\subsection{Pure neutron matter}

\begin{figure}[tb]
\begin{center}
\includegraphics[angle=0,width=8cm]{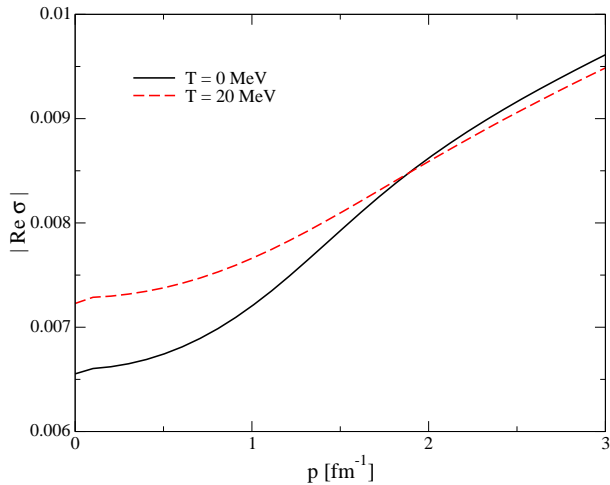}
\caption{
(Color online) Dependence of the one-pion exchange contribution to the nucleon 
self-energy in pure neutron matter on the momentum, at density $n = 0.1$ fm$^{-3}$. 
The solid line shows the real part of the self-energy 
for temperature $T=0$ while the dashed line is for $T=20$ MeV.
}
\label{fig:self_neutron}
\end{center}
\end{figure}
The reduced self-energy of neutrons in pure neutron matter is shown in
Fig.~\ref{fig:self_neutron}. Compared to isospin-symmetric nuclear 
matter, the self-energy in neutron matter is
smaller. This is due to the fact that in pure neutron matter  one-pion
exchange involves only the $\pi^0$-meson. 
At non-zero temperature, as in the case of isospin-symmetric
nuclear matter, we observe that, for low momenta, the contribution of the
self-energy  is larger, but the difference among the 
two cases tends to disappear with increasing momentum.

\section{Conclusions}
\label{sec:conclusions}

In this work we have combined the methods of TFT and chiral
Lagrangians to compute the self-energy of nucleons to leading order in
the chiral expansion. In doing so, we have maintained the covariance
of the pion and nucleon propagators and we have imposed
self-consistency by solving a Schwinger-Dyson equation for the nucleon
self-energy.  Our approach has also been applied to pure neutron
matter, with similar results.

Clearly, to obtain a consistent phenomenology of both
isospin-symmetric nuclear and pure neutron matter, one needs to
introduce contact interactions which account for the short-range
two-body and three-body interactions. Furthermore, the importance of
second-order pion exchange was already stressed by Lutz et
al.~\cite{2000PhLB..474....7L} and KFW. In this respect further steps
might be undertaken to resolve the relativistic dynamics of pions by
including higher-order terms in the chiral expansion and incorporating
$\Delta$-isobar excitations.

Methodologically, our approach differs from similar works since we
address the nucleon self-energy within a chiral effective thermal field theory
by keeping a relativistic framework, and at the same time we impose
self-consistency by solving a Schwinger-Dyson equation.

The in-medium electromagnetic and weak interactions of nucleons can be
computed in a relativistically covariant manner starting from the
self-consistent propagators derived above.

\section*{Acknowledgments}
This work was partially supported by the HGS-HIRe graduate program at
Frankfurt University (G. C.). We thank E. S.  Fraga, R. D. Pisarski,
and J. Schaffner-Bielich for discussions,  S. Gandolfi for
correspondence and the anonymous referee for pointing out a sign error
in an earlier version. Furthermore, we are indebted to Bengt Friman for his
continuous interest in this work and useful suggestions.


%
\end{document}